\documentclass[preprint]{aastex63}

\usepackage{graphicx}
\usepackage{bm}
\usepackage{amsmath}
\usepackage{mathtools}
\usepackage{epsf}
\usepackage[figuresright]{rotating}
\usepackage{mathrsfs}

\shorttitle{Change ratios of magnetic helicity and free energy during major solar flares}
\shortauthors{Wang et al.}

\begin{document}

\title{Change ratios of magnetic helicity and magnetic free energy during major solar flares}

\correspondingauthor{Quan Wang}
\email{wangquan@nao.cas.cn}

\author{Quan Wang}
\affiliation{Key Laboratory of Solar Activity, National Astronomical Observatories, Chinese Academy of Sciences, 100101 Beijing, China}
\affiliation{School of Astronomy and Space Sciences, University of Chinese Academy of Sciences, 100049 Beijing, China}

\author{Mei Zhang}
\affiliation{Key Laboratory of Solar Activity, National Astronomical Observatories, Chinese Academy of Sciences, 100101 Beijing, China}
\affiliation{School of Astronomy and Space Sciences, University of Chinese Academy of Sciences, 100049 Beijing, China}

\author{Shangbin Yang}
\affiliation{Key Laboratory of Solar Activity, National Astronomical Observatories, Chinese Academy of Sciences, 100101 Beijing, China}
\affiliation{School of Astronomy and Space Sciences, University of Chinese Academy of Sciences, 100049 Beijing, China}

\author{Xiao Yang}
\affiliation{Key Laboratory of Solar Activity, National Astronomical Observatories, Chinese Academy of Sciences, 100101 Beijing, China}

\author{Xiaoshuai Zhu}
\affiliation{State Key Laboratory of Space Weather, National Space Science Center, Chinese Academy of Sciences, 100190 Beijing, China}

\begin{abstract}

Magnetic helicity is an important concept in solar physics, with a number of theoretical statements pointing out the important role of magnetic helicity in solar flares and coronal mass ejections (CMEs). Here we construct a sample of 47 solar flares, which contains 18 no-CME-associated confined flares and 29 CME-associated eruptive flares. We calculate the change ratios of magnetic helicity and magnetic free energy before and after these 47 flares. Our calculations show that the change ratios of magnetic helicity and magnetic free energy show distinct different distributions in confined flares and eruptive flares. The median value of the change ratios of magnetic helicity in confined flares is $-0.8$\%, while this number is $-14.5$\% for eruptive flares. For the magnetic free energy, the median value of the change ratios is  $-4.3$\% for confined flares, whereas this number is $-14.6$\% for eruptive flares. This statistical result, using observational data, is well consistent with the theoretical understandings that magnetic helicity is approximately conserved in the magnetic reconnection, as shown by confined flares, and the CMEs take away magnetic helicity from the corona, as shown by eruptive flares.

\end{abstract}

\keywords{Sun: Magnetic fields -- Sun: Corona -- Sun: Flares}

\section{Introduction}
\label{sect:intro}

Magnetic helicity is a physical quantity that describes the topology and complexity of a magnetic field \citep{Elsasser1956, Woltjer1958, Moffatt1969}. It is believed that magnetic helicity, as a conservative physical quantity, plays an important role in MHD system \citep{Taylor1974} and in solar activity \citep{Zhang2005}.

\citet{Berger1984a} gave a rigorous upper limit of the amount of magnetic helicity that could be dissipated through the constant mean resistivity in resistive MHD and proved that the typical magnetic helicity dissipation time scale in the corona is far larger than the magnetic energy dissipation time scale. This hypothesis has led to a series of research on estimating the magnetic helicity and energy dissipation in the corona, \citet{Yang2013, Yang2018} and \citet{Pariat2015} for example. However, these studies are mainly based on numerical simulation data.

The conservation principle of magnetic helicity also led to a theoretical hypothesis that, the accumulated magnetic helicity in the corona will exceed the upper bound of a force-free magnetic field and result in a non-equilibrium situation such as a coronal mass ejection (CME) \citep{Zhang2006, Zhang2008, Zhang2012} and the resultant CME will carry the excess magnetic helicity out of the corona and spread them into the interplanetary space \citep{Rust1994, Low1996}.

Many studies have then investigated the occurrence of flares and CMEs, with the particular attention on their relationship with coronal magnetic helicity and magnetic free energy. \citet{Tziotziou2012} found that active regions (ARs) associated with eruptive flares appear well segregated from confined, in both relative magnetic helicity and magnetic free energy. The major (at least M-class) flares are trend to be occurring in ARs with relative magnetic helicity and magnetic free energy exceeding $2 \times 10^{42}$ Mx$^2$ and $4 \times 10^{31}$ erg, respectively. \citet{Liokati2022} targeted this issue too. By studying magnetic helicity and magnetic energy injections in 52 emerging ARs, they gave the magnetic helicity and magnetic energy thresholds for eruption as of $9 \times 10^{41}$ Mx$^2$ and $2 \times 10^{32}$ erg, respectively. \citet{Pariat2017} took an approach from another point of view. They proposed that the relationship between the two parts of helicity, that is, the ratio between the magnetic helicity of solenoidal component of the current-carrying magnetic field ($|H_J|$) and the mutual helicity between potential field and current-carrying field ($|H_\nu|$), could be used as a criterion of the trigger of eruptive flares. \citet{Thalmann2019b} checked this statement using two ARs, one (AR 11158) prolific in eruptive flares and one (AR 12192) in confined flares. They confirmed that $|H_J|/$$|H_\nu|$ shows ``a strong ability to indicate the eruptive potential of a solar AR''. But they also pointed out that ``$|H_J|/$$|H_\nu|$ does not seem to be indicative for the magnitude or type of an upcoming flare (confined or eruptive)''.

Above studies are mainly concerned with the pre-event conditions for CME eruptions. There seems not much studies in discussing the changes of magnetic helicity and magnetic free energy during the solar flares, particularly using observational data. Using MDI/SOHO data, \citet{Valori2015} studied the magnetic helicity flux through the photosphere in NOAA 10365. They used a linear force-free field approximation to translate the obtained magnitudes of accumulated helicity into a series of force-free parameters $\alpha$. Based on these force-free parameters they have estimated the free magnetic energy in the corona and found that ``the higher the value of the accumulated coronal helicity, the smaller the force-free parameter variation required to produce the same decrease in the free energy during the CMEs''.  It is only recently that \citet{Liu2023} used the non-linear force-free field (NLFFF) extrapolation to study the changes of magnetic energy and helicity in solar active regions using a sample of 21 X-class flares. Their study gave a number of interesting results. However, they used a superposed epoch analysis approach where  all eruptive or confined profiles are averaged to get the corresponding time evolution profiles. 

In this paper, we intend to discuss the time evolution of solar flares, with a particular attention to check whether there is a systematic difference on the changes of magnetic helicity and free energy between CME-associated eruptive flares and no-CME-associated confined flares. We constructed a sample of 47 solar flares in solar cycle 24, which contains 18  confined flares and 29  eruptive flares. We calculated the change ratios of coronal magnetic helicity and magnetic free energy before and after these 47 flares. Different from \citet{Liu2023}, we did not used the superposed epoch analysis approach. We calculated the specific values of magnetic helicity and magnetic free energy throughout a series of time evolution for each flare. Also our sample is larger than theirs. The rest of the paper is organized as follows. Section 2 describes the sample, data and methods we use. Section 3 presents the analysis and the results. A brief summary and discussion is given in Section 4.

\section{The Sample, Data and Methods}
\label{sect:Data}

\subsection{The Sample}

Our sample consists of 47 major solar flares. These flares are selected according to following criterions.  a) The X-ray class of the flare is above M4.0.  b) The latitude of the flare is between 30 degrees north and 30 degrees south, and its longitude is within 45 degrees from the central meridian.  c) No other flares above M4.0 occur within two hours before and after the flare.  The information of these flares, such as the location on the heliocentric coordinate, the GOES soft X-ray class, the start, peak and end time of the flare ($T_{start}$, $T_{peak}$ and $T_{end}$), taken from the NOAA solar event report\footnote{ftp://ftp.ngdc.noaa.gov/STP/swpc\_products/daily\_reports/solar\_event\_reports/}, are listed in Tables \ref{Tab1} and \ref{Tab2}. These selected 47 flare events are from 23 active regions. Out of these 47 flares, 31 (66.0\%) are located in the southern hemisphere and 16 (34.0\%) in the northern hemisphere.

\begin{sidewaystable}  \tiny \addtolength{\tabcolsep}{-4pt}
\vspace*{1mm}
\begin{center}
\caption{Information and results of the 18 confined flare events}
\label{Tab1}
\renewcommand\arraystretch{1.0}
 \begin{tabular}{ccccccccccrrrrrrrcrr}
  \hline\noalign{\smallskip}
\scriptsize{No.}  & \scriptsize{Date}  & \scriptsize{$T_{start}$}  & \scriptsize{$T_{peak}$} & \scriptsize{$T_{end}$}  & \scriptsize{Class}  & \scriptsize{NOAA}  & \scriptsize{POS}  & \scriptsize{Box Size}  &  \scriptsize{Flux}  &  \scriptsize{$H_R^0$} & \scriptsize{$H_R^1$}  &  \scriptsize{d$H_R$}  &  \scriptsize{$\eta_H$}  &  \scriptsize{$\eta_H^{\prime}$}  &  \scriptsize{$E_f^0$} & \scriptsize{$E_f^1$} &  \scriptsize{d$E_m$}  & \scriptsize{$\eta_E$} & \scriptsize{$\eta_E^{\prime}$}  \\  
  &   &   &  &   &   &   &   &  \tiny{($arcsec^3$)} &  \tiny{($10^{20}$ Mx)}  &  \tiny{($10^{42}$ Mx$^2$)} & \tiny{($10^{42}$ Mx$^2$)} & \tiny{($10^{41}$ Mx$^2$)} &  \tiny{(\%)} & \tiny{(\%)} &  \tiny{($10^{31}$ erg)} & \tiny{($10^{31}$ erg)} & \tiny{($10^{31}$ erg)} &  \tiny{(\%)} & \tiny{(\%)} \\  
\noalign{\smallskip}\hline\noalign{\smallskip}
01 & 20120509 & 12:21 & 12:32 & 12:36 & M4.7 & 11476 & N13E31 & 400$\times$200$\times$200 & 319  & $-5.55$ $\pm$ 0.11   & $-5.54$ $\pm$ 0.05   & 0.52     & $-0.18$  & 0.77     & 13.5 $\pm$ 0.1  & 11.2 $\pm$ 0.4  & 0.43 & $-17.00$ & $-20.16$ \\
02 & 20120705 & 03:25 & 03:36 & 03:39 & M4.7 & 11515 & S17W23 & 400$\times$200$\times$200 & 289  & $-14.04$ $\pm$ 0.13  & $-14.58$ $\pm$ 0.43  & $-1.11$  & 3.89     & 3.10     & 23.3 $\pm$ 0.2  & 24.7 $\pm$ 0.7  & 0.41 & 5.82     & 4.04     \\
03 & 20140202 & 09:24 & 09:31 & 09:53 & M4.4 & 11967 & S11E13 & 400$\times$200$\times$200 & 400  & $-18.87$ $\pm$ 0.18  & $-18.34$ $\pm$ 0.09  & $-0.41$  & $-2.80$  & $-3.02$  & 48.6 $\pm$ 1.3  & 46.5 $\pm$ 1.8  & 1.32 & $-4.31$  & $-7.03$  \\
04 & 20140204 & 03:57 & 04:00 & 04:06 & M5.2 & 11967 & S14W06 & 400$\times$200$\times$200 & 401  & $-26.13$ $\pm$ 0.05  & $-26.32$ $\pm$ 0.01  & $-0.43$  & 0.71     & 0.55     & 49.1 $\pm$ 0.2  & 46.2 $\pm$ 0.2  & 0.37 & $-5.92$  & $-6.67$  \\
05 & 20141020 & 16:00 & 16:37 & 16:55 & M4.5 & 12192 & S14E37 & 512$\times$256$\times$256 & 770  & $-263.52$ $\pm$ 0.97 & $-264.43$ $\pm$ 0.97 & $-16.38$ & 0.35     & $-0.27$  & 74.4 $\pm$ 1.2  & 72.1 $\pm$ 2.0  & 5.18 & $-3.07$  & $-10.03$ \\
06 & 20141022 & 01:16 & 01:59 & 02:28 & M8.7 & 12192 & S13E21 & 512$\times$256$\times$256 & 760  & $-279.02$ $\pm$ 0.66 & $-264.95$ $\pm$ 1.42 & $-16.46$ & $-5.04$  & $-5.63$  & 124.3 $\pm$ 0.5 & 104.5 $\pm$ 1.6 & 8.56 & $-15.89$ & $-22.78$ \\
07 & 20141022 & 14:02 & 14:28 & 14:50 & X1.6 & 12192 & S14E13 & 512$\times$256$\times$256 & 735  & $-260.18$ $\pm$ 0.55 & $-258.21$ $\pm$ 2.07 & $-13.30$ & $-0.76$  & $-1.27$  & 93.8 $\pm$ 0.5  & 94.6 $\pm$ 2.1  & 6.61 & 0.93     & $-6.12$  \\
08 & 20141024 & 21:07 & 21:41 & 22:13 & X3.1 & 12192 & S16W21 & 512$\times$256$\times$256 & 820  & $-371.36$ $\pm$ 0.14 & $-355.13$ $\pm$ 0.88 & $-25.43$ & $-4.37$  & $-5.06$  & 162.6 $\pm$ 1.2 & 139.8 $\pm$ 0.8 & 8.84 & $-14.00$ & $-19.43$ \\
09 & 20141025 & 16:55 & 17:08 & 18:11 & X1.0 & 12192 & S16W31 & 512$\times$256$\times$256 & 890  & $-378.00$ $\pm$ 1.33 & $-359.95$ $\pm$ 1.38 & $-14.35$ & $-4.77$  & $-5.15$  & 152.9 $\pm$ 1.2 & 137.1 $\pm$ 0.8 & 5.53 & $-10.35$ & $-13.96$ \\
10 & 20141026 & 10:44 & 10:56 & 11:18 & X2.0 & 12192 & S18W40 & 512$\times$256$\times$256 & 934  & $-313.79$ $\pm$ 2.62 & $-302.26$ $\pm$ 1.65 & $-6.29$  & $-3.67$  & $-3.88$  & 97.6 $\pm$ 3.1  & 85.9 $\pm$ 2.6  & 3.36 & $-12.01$ & $-15.45$ \\
11 & 20141026 & 18:07 & 18:15 & 18:20 & M4.2 & 12192 & S16W34 & 512$\times$256$\times$256 & 988  & $-304.65$ $\pm$ 0.62 & $-301.18$ $\pm$ 0.23 & $-2.64$  & $-1.14$  & $-1.22$  & 92.8 $\pm$ 0.3  & 90.8 $\pm$ 1.0  & 1.19 & $-2.13$  & $-3.41$  \\
12 & 20141027 & 00:06 & 00:34 & 00:44 & M7.1 & 12192 & S14W44 & 512$\times$256$\times$256 & 1007 & $-298.74$ $\pm$ 1.97 & $-304.10$ $\pm$ 3.06 & $-6.01$  & 1.79     & 1.59     & 97.2 $\pm$ 1.9  & 101.6 $\pm$ 2.0 & 4.22 & 4.51     & 0.17     \\
13 & 20141204 & 18:05 & 18:25 & 18:56 & M6.1 & 12222 & S20W31 & 400$\times$200$\times$200 & 252  & $-10.52$ $\pm$ 0.02  & $-9.93$ $\pm$ 0.03   & $-2.75$  & $-5.61$  & $-8.23$  & 12.0 $\pm$ 0.1  & 10.8 $\pm$ 0.1  & 0.80 & $-10.58$ & $-17.26$ \\
14 & 20150312 & 13:50 & 14:08 & 14:13 & M4.2 & 12297 & S15E06 & 400$\times$200$\times$200 & 169  & 10.15 $\pm$ 0.04     & 10.22 $\pm$ 0.15     & 0.72     & 0.72     & 0.01     & 23.1 $\pm$ 0.2  & 22.5 $\pm$ 0.2  & 0.42 & $-2.66$  & $-4.48$  \\
15 & 20150824 & 07:26 & 07:33 & 07:35 & M5.6 & 12403 & S15W04 & 400$\times$200$\times$200 & 322  & 12.69 $\pm$ 0.12     & 12.72 $\pm$ 0.05     & 1.03     & 0.21     & $-0.60$  & 24.6 $\pm$ 0.2  & 24.8 $\pm$ 0.1  & 0.35 & 0.54     & $-0.88$  \\
16 & 20150928 & 14:53 & 14:58 & 15:03 & M7.6 & 12422 & S20W28 & 360$\times$180$\times$180 & 255  & $-9.37$ $\pm$ 0.08   & $-8.46$ $\pm$ 0.07   & $-0.20$  & $-9.76$  & $-9.98$  & 14.3 $\pm$ 0.1  & 13.0 $\pm$ 0.2  & 0.36 & $-9.21$  & $-11.75$ \\
17 & 20170905 & 01:03 & 01:08 & 01:11 & M4.2 & 12673 & S09W14 & 300$\times$200$\times$200 & 198  & $-25.96$ $\pm$ 0.07  & $-26.09$ $\pm$ 0.10  & $-0.74$  & 0.52     & 0.23     & 45.0 $\pm$ 0.2  & 44.4 $\pm$ 0.1  & 0.26 & $-1.27$  & $-1.84$  \\
18 & 20170906 & 08:57 & 09:10 & 09:17 & X2.2 & 12673 & S08W32 & 300$\times$200$\times$200 & 266  & $-39.12$ $\pm$ 1.18  & $-30.42$ $\pm$ 0.83  & $-1.54$  & $-22.25$ & $-22.64$ & 71.5 $\pm$ 2.3  & 54.7 $\pm$ 0.5  & 0.55 & $-23.43$ & $-24.19$ \\
  \noalign{\smallskip}\hline
\end{tabular}
\end{center}
\end{sidewaystable}

\begin{sidewaystable} \tiny \addtolength{\tabcolsep}{-4pt}
\vspace*{1mm}
\begin{center}
\caption{Information and results of the 29 eruptive flare events}
\label{Tab2}
 \begin{tabular}{ccccccccccrrrrrrrcrr}
  \hline\noalign{\smallskip}
\scriptsize{No.}  & \scriptsize{Date}  & \scriptsize{$T_{start}$}  & \scriptsize{$T_{peak}$} & \scriptsize{$T_{end}$}  & \scriptsize{Class}  & \scriptsize{NOAA}  & \scriptsize{POS}  & \scriptsize{Box Size}  &  \scriptsize{Flux}  &  \scriptsize{$H_R^0$} & \scriptsize{$H_R^1$}  &  \scriptsize{d$H_R$}  &  \scriptsize{$\eta_H$}  &  \scriptsize{$\eta_H^{\prime}$}  &  \scriptsize{$E_f^0$} & \scriptsize{$E_f^1$} &  \scriptsize{d$E_m$}  & \scriptsize{$\eta_E$} & \scriptsize{$\eta_E^{\prime}$}  \\  
  &   &   &  &   &   &   &   &  \tiny{($arcsec^3$)} &  \tiny{($10^{20}$ Mx)}  &  \tiny{($10^{42}$ Mx$^2$)} & \tiny{($10^{42}$ Mx$^2$)} & \tiny{($10^{41}$ Mx$^2$)} &  \tiny{(\%)} & \tiny{(\%)} &  \tiny{($10^{31}$ erg)} & \tiny{($10^{31}$ erg)} & \tiny{($10^{31}$ erg)} &  \tiny{(\%)} & \tiny{(\%)} \\  
\noalign{\smallskip}\hline\noalign{\smallskip}
19 & 20110213 & 17:28 & 17:38 & 17:47 & M6.6 & 11158 & S20E04 & 360$\times$180$\times$180 & 116 & 1.96 $\pm$ 0.03     & 1.60 $\pm$ 0.01     & 0.32    & $-18.39$ & $-20.05$ & 11.2 $\pm$ 0.01 & 9.6 $\pm$ 0.1   & 0.31 & $-14.55$ & $-17.33$ \\
20 & 20110215 & 01:44 & 01:56 & 02:06 & X2.2 & 11158 & S20W12 & 360$\times$180$\times$180 & 155 & 6.02 $\pm$ 0.09     & 4.60 $\pm$ 0.01     & 0.32    & $-23.70$ & $-24.23$ & 23.1 $\pm$ 0.02 & 19.5 $\pm$ 0.3  & 0.51 & $-15.62$ & $-17.82$ \\
21 & 20110803 & 13:17 & 13:48 & 14:10 & M6.0 & 11261 & N16W30 & 400$\times$200$\times$200 & 168 & 7.10 $\pm$ 0.02     & 7.12 $\pm$ 0.06     & $-0.14$ & 0.37     & 0.58     & 17.2 $\pm$ 0.2  & 14.5 $\pm$ 0.02 & 0.18 & $-15.58$ & $-16.63$ \\
22 & 20110804 & 03:41 & 03:57 & 04:04 & M9.3 & 11261 & N16W38 & 400$\times$200$\times$200 & 191 & 8.40 $\pm$ 0.06     & 6.85 $\pm$ 0.07     & 0.04    & $-18.51$ & $-18.56$ & 17.8 $\pm$ 0.5  & 13.6 $\pm$ 0.4  & 0.16 & $-24.04$ & $-24.92$ \\
23 & 20110906 & 01:35 & 01:50 & 02:05 & M5.3 & 11283 & N14W07 & 400$\times$200$\times$200 & 144 & 3.11 $\pm$ 0.08     & 4.01 $\pm$ 0.12     & $-0.12$ & 28.81    & 29.18    & 5.1 $\pm$ 0.1   & 5.2 $\pm$ 0.1   & 0.19 & 2.49     & $-1.17$  \\
24 & 20110906 & 22:12 & 22:20 & 22:24 & X2.1 & 11283 & N14W18 & 400$\times$200$\times$200 & 142 & 2.89 $\pm$ 0.10     & 2.27 $\pm$ 0.03     & $-0.08$ & $-21.70$ & $-21.41$ & 7.7 $\pm$ 0.1   & 5.0 $\pm$ 0.04  & 0.08 & $-35.68$ & $-36.70$ \\
25 & 20110907 & 22:32 & 22:38 & 22:44 & X1.8 & 11283 & N14W28 & 400$\times$200$\times$200 & 157 & 4.39 $\pm$ 0.05     & 3.76 $\pm$ 0.12     & $-0.01$ & $-14.50$ & $-14.47$ & 8.5 $\pm$ 0.1   & 3.8 $\pm$ 0.1   & 0.09 & $-55.30$ & $-56.40$ \\
26 & 20120123 & 03:38 & 03:59 & 04:34 & M8.7 & 11402 & N28W21 & 400$\times$200$\times$200 & 242 & $-7.47$ $\pm$ 0.10  & $-12.75$ $\pm$ 0.14 & $-0.59$ & 70.74    & 69.95    & 15.3 $\pm$ 0.5  & 20.7 $\pm$ 0.1  & 0.98 & 35.48    & 29.09    \\
27 & 20120307 & 00:02 & 00:24 & 00:40 & X5.4 & 11429 & N17E27 & 320$\times$160$\times$160 & 268 & $-59.53$ $\pm$ 0.06 & $-44.01$ $\pm$ 0.32 & $-2.00$ & $-26.07$ & $-26.40$ & 90.5 $\pm$ 0.1  & 63.7 $\pm$ 1.0  & 0.50 & $-29.63$ & $-30.19$ \\
28 & 20120309 & 03:22 & 03:53 & 04:18 & M6.3 & 11429 & N17W03 & 320$\times$160$\times$160 & 236 & $-30.76$ $\pm$ 0.09 & $-30.26$ $\pm$ 0.11 & $-2.14$ & $-1.61$  & $-2.31$  & 60.9 $\pm$ 0.1  & 60.3 $\pm$ 0.2  & 1.33 & $-1.10$  & $-3.28$  \\
29 & 20120310 & 17:15 & 17:44 & 18:30 & M8.4 & 11429 & N17W24 & 320$\times$160$\times$160 & 233 & $-28.05$ $\pm$ 0.14 & $-26.61$ $\pm$ 0.22 & $-2.05$ & $-5.13$  & $-5.86$  & 61.4 $\pm$ 0.2  & 53.7 $\pm$ 0.7  & 1.61 & $-12.56$ & $-15.18$ \\
30 & 20120307 & 01:05 & 01:14 & 01:23 & X1.3 & 11430 & N17E27 & 200$\times$200$\times$200 & 67  & $-2.77$ $\pm$ 0.01  & $-2.73$ $\pm$ 0.08  & $-0.06$ & $-1.45$  & $-1.67$  & 6.8 $\pm$ 0.1   & 6.3 $\pm$ 0.1   & 0.05 & $-6.87$  & $-7.57$  \\
31 & 20120702 & 10:43 & 10:52 & 11:12 & M5.6 & 11515 & S17E08 & 400$\times$200$\times$200 & 194 & $-14.32$ $\pm$ 0.13 & $-14.01$ $\pm$ 0.13 & $-1.06$ & $-2.11$  & $-2.86$  & 12.7 $\pm$ 0.03 & 12.3 $\pm$ 0.2  & 0.51 & $-3.47$  & $-7.51$  \\
32 & 20120704 & 09:47 & 09:55 & 10:16 & M5.3 & 11515 & S20W18 & 400$\times$200$\times$200 & 244 & $-13.60$ $\pm$ 0.05 & $-13.01$ $\pm$ 0.13 & $-1.77$ & $-4.36$  & $-5.66$  & 18.5 $\pm$ 0.2  & 16.9 $\pm$ 0.4  & 0.57 & $-8.84$  & $-11.92$ \\
33 & 20130411 & 06:55 & 07:16 & 07:29 & M6.5 & 11719 & N09W12 & 400$\times$200$\times$200 & 126 & $-0.79$ $\pm$ 0.01  & $-1.14$ $\pm$ 0.04  & $-0.03$ & 44.78    & 44.38    & 4.3 $\pm$ 0.1   & 4.3 $\pm$ 0.04  & 0.26 & 0.16     & $-5.93$  \\
34 & 20131024 & 00:21 & 00:30 & 00:50 & M9.3 & 11877 & S10E08 & 256$\times$128$\times$128 & 106 & 6.08 $\pm$ 0.03     & 5.92 $\pm$ 0.02     & 0.26    & $-2.68$  & $-3.11$  & 5.7 $\pm$ 0.04  & 5.6 $\pm$ 0.03  & 0.53 & $-0.57$  & $-9.83$  \\
35 & 20131105 & 22:07 & 22:12 & 22:15 & X3.3 & 11890 & S13E44 & 400$\times$200$\times$200 & 367 & 5.61 $\pm$ 0.03     & 3.24 $\pm$ 0.23     & 0.08    & $-42.19$ & $-42.33$ & 29.2 $\pm$ 0.2  & 21.3 $\pm$ 0.2  & 0.17 & $-26.96$ & $-27.54$ \\
36 & 20131108 & 04:20 & 04:26 & 04:29 & X1.1 & 11890 & S14E15 & 400$\times$200$\times$200 & 251 & 6.50 $\pm$ 0.26     & 5.44 $\pm$ 0.03     & 0.22    & $-16.44$ & $-16.79$ & 16.0 $\pm$ 0.2  & 13.2 $\pm$ 0.2  & 0.25 & $-17.53$ & $-19.09$ \\
37 & 20131110 & 05:08 & 05:14 & 05:18 & X1.1 & 11890 & S14W13 & 400$\times$200$\times$200 & 236 & 5.74 $\pm$ 0.13     & 4.26 $\pm$ 0.07     & 0.04    & $-25.75$ & $-25.83$ & 14.6 $\pm$ 0.1  & 12.8 $\pm$ 0.1  & 0.19 & $-12.62$ & $-13.92$ \\
38 & 20140329 & 17:35 & 17:48 & 17:54 & X1.0 & 12017 & N11W32 & 200$\times$100$\times$100 & 53  & 1.82 $\pm$ 0.02     & 1.18 $\pm$ 0.01     & 0.14    & $-35.44$ & $-36.21$ & 7.7 $\pm$ 0.1   & 3.7 $\pm$ 0.1   & 0.05 & $-51.52$ & $-52.17$ \\
39 & 20140418 & 12:31 & 13:03 & 13:20 & M7.3 & 12036 & S20W34 & 256$\times$128$\times$128 & 119 & 2.41 $\pm$ 0.14     & 1.38 $\pm$ 0.09     & 0.19    & $-42.48$ & $-43.28$ & 11.6 $\pm$ 0.4  & 8.9 $\pm$ 0.1   & 0.47 & $-22.66$ & $-26.75$ \\
40 & 20141107 & 16:53 & 17:26 & 17:34 & X1.6 & 12205 & N17E40 & 400$\times$200$\times$200 & 250 & 4.88 $\pm$ 0.11     & 0.68 $\pm$ 0.24     & 0.52    & $-86.10$ & $-87.17$ & 42.0 $\pm$ 0.8  & 22.3 $\pm$ 0.3  & 0.49 & $-46.93$ & $-48.10$ \\
41 & 20141218 & 21:41 & 21:58 & 22:25 & M6.9 & 12241 & S11E10 & 320$\times$160$\times$160 & 187 & 14.81 $\pm$ 0.15    & 13.22 $\pm$ 0.06    & 0.49    & $-10.76$ & $-11.09$ & 20.9 $\pm$ 0.5  & 18.9 $\pm$ 0.1  & 1.22 & $-9.49$  & $-15.32$ \\
42 & 20150310 & 03:19 & 03:24 & 03:28 & M5.1 & 12297 & S15E40 & 400$\times$200$\times$200 & 224 & 22.23 $\pm$ 0.43    & 20.82 $\pm$ 0.42    & 0.23    & $-6.35$  & $-6.45$  & 25.8 $\pm$ 0.5  & 23.7 $\pm$ 0.7  & 0.09 & $-8.19$  & $-8.53$  \\
43 & 20150311 & 16:11 & 16:22 & 16:29 & X2.1 & 12297 & S17E21 & 400$\times$200$\times$200 & 178 & 6.45 $\pm$ 0.04     & 9.60 $\pm$ 0.10     & 0.05    & 48.62    & 48.55    & 13.5 $\pm$ 0.1  & 19.1 $\pm$ 0.1  & 0.21 & 41.18    & 39.60    \\
44 & 20150622 & 17:39 & 18:23 & 18:51 & M6.5 & 12371 & N12W08 & 400$\times$200$\times$200 & 241 & $-50.76$ $\pm$ 0.18 & $-42.01$ $\pm$ 0.14 & 0.32    & $-17.22$ & $-17.16$ & 49.1 $\pm$ 0.3  & 36.3 $\pm$ 0.1  & 1.84 & $-26.04$ & $-29.79$ \\
45 & 20150625 & 08:02 & 08:16 & 09:05 & M7.9 & 12371 & N09W42 & 400$\times$200$\times$200 & 300 & $-47.88$ $\pm$ 0.45 & $-29.99$ $\pm$ 0.19 & $-1.04$ & $-37.36$ & $-37.58$ & 46.5 $\pm$ 0.5  & 35.5 $\pm$ 0.2  & 1.80 & $-23.69$ & $-27.57$ \\
46 & 20170904 & 20:28 & 20:33 & 20:37 & X5.5 & 12673 & S10W11 & 300$\times$200$\times$200 & 188 & $-24.87$ $\pm$ 0.10 & $-24.80$ $\pm$ 0.14 & $-0.77$ & $-0.30$  & $-0.61$  & 43.2 $\pm$ 0.4  & 40.7 $\pm$ 0.5  & 0.31 & $-5.81$  & $-6.53$  \\
47 & 20170906 & 11:53 & 12:02 & 12:10 & X9.3 & 12673 & S08W34 & 300$\times$200$\times$200 & 264 & $-30.16$ $\pm$ 1.10 & $-20.26$ $\pm$ 0.84 & $-1.64$ & $-32.83$ & $-33.38$ & 54.3 $\pm$ 1.7  & 34.3 $\pm$ 1.1  & 0.77 & $-36.75$ & $-38.17$ \\
  \noalign{\smallskip}\hline
\end{tabular}
\end{center}
\end{sidewaystable}

Flares associated with CMEs are usually referred to as ``eruptive'' events, while flares that are not accompanied by CMEs are called ``confined'' events \citep{Moore2001}. We use the information of CMEs in the Coordinated Data Analysis Workshops (CDAW) CME catalogue\footnote{\url{https://cdaw.gsfc.nasa.gov/CME_list/}} to determine whether a flare event is associated with a CME or not. We regard a flare event as ``eruptive'' if the following two criteria are satisfied: a) the time difference between the onset time of a CME and the peak time of the flare is less than two hours; b) the position angle difference between the CME and the flare is less than half of the CME angular width \citep{Joshi2018, Zhang2021}. We have also verified our identification with previous works \citep{Kazachenko2017, Li2020}. Out of these 47 flares, 18 (38.3\%) events are ``confined'', information of which are listed in Table \ref{Tab1}, and 29 (61.7\%) events are ``eruptive'', information of which are listed in Table \ref{Tab2}.

\subsection{The Data}

The magnetograms we use are taken by Helioseismic and Magnetic Imager (HMI)/SDO. HMI \citep{Scherrer2012} observes the full solar disk at Fe\,{\small I} $\lambda$6173 with a $4096 \times 4096$ CCD detector to study the oscillations and the magnetic fields on the solar photosphere. The spatial resolution is 0.91$''$ with a 0$''$.5 pixel size. The vector magnetograms are obtained, using a Milne–Eddington based inversion code \citep{Borrero2011}, from the filtergrams taken at six wavelength positions. The azimuthal $180^{\circ}$ ambiguity was resolved by the ``minimum energy'' algorithm \citep{Metcalf2006}. 

This study uses the hmi.sharp\_cea\_720s series of active-region vector magnetograms. In this series of the data, the disambiguated vector magnetograms are deprojected using Lambert cylindrical equal area projection method, presented as ($B_r$, $B_{\theta}$, $B_{\phi}$) in heliocentric spherical coordinates which corresponds to ($B_z$, -$B_y$, $B_x$) in heliographic coordinates. After downloading these data, we cut each magnetogram for a better use of doing extrapolation. The field of views (FOVs) of these magnetograms can be found in Tables \ref{Tab1} and \ref{Tab2}, as the first two numbers in the ``box size'' column.

\subsection{The Extrapolation Method}

Since so far we still have no accurate coronal magnetic field measurements on a  daily basis, it has become a common approach in the solar physics community to reconstruct the 3D coronal magnetic field based on nonlinear force-free field (NLFFF) assumption \citep{Wheatland2000, Wiegelmann2008, Wiegelmann2017, Wiegelmann2021}, using the observed photospheric magnetogram as the boundary condition.

In this paper we adopt the weighted optimization NLFFF extrapolation method developed in \citet{Wiegelmann2004} to reconstruct the coronal magnetic field. Since the observed photospheric magnetic field is usually not strictly force-free \citep{ZhangXM2017},  the observed magnetograms need to be revised toward suitable boundary conditions for NLFFF extrapolation. All vector magnetograms used in this study are preprocessed following an algorithm proposed in \citet{Wiegelmann2006}. The specific process of this algorithm were described in \citet{Wiegelmann2006} and \citet{Schrijver2006}. In this study, we have used the most updated version of the code. The basic code of the extrapolation is not changed in the new version, however, a new multigrid extension of the code has been implemented. It is found that the multigrid version of the code converges to a lower joint measure L value \citep{Wheatland2000,Schrijver2006}, which means a higher quality of the extrapolation is achieved.  Also it makes the code run more efficiently. In this work, we use this multigrid version in all of the extrapolations.

The sizes of the extrapolation box of each flare are listed in Tables 1 and 2. Note that unlike many other studies, we have used the original spatial resolution of HMI/SDO data. This makes our calculations quite time-comsuming. But it probably worths of it, as the estimation of the two controlling parameters shows that our extrapolation fields are of good quality, a point we will come back later in the next section. 

\subsection{The Calculation of Relative Magnetic Helicity and Magnetic Free Energy}

The magnetic helicity we refer to and calculate in this paper are all ``relative magnetic helicity'', a concept proposed by \citet{Berger1984b}, which defines the magnetic helicity in open and multiply connected volumes such as the solar corona. This physical quantity is given by the Finn–Antonsen formula \citep{Finn1985},
\begin{equation}
    \textsl{H}_{R}=  \int (\bm{A} + \bm{A}_p) \cdot (\bm{B} - \bm{B}_p) d^{3}x \ ,
\label{equ1}
\end{equation}
where $\bm{A}$ is the vector potential of the studied magnetic field $\bm{B}$ and $\bm{A}_p$ is the vector potential of the current-free potential field $\bm{B}_p$, which is uniquely determined by the normal component of the magnetic field at the boundaries.

The magnetic free energy we studied is defined as $E_f = E_m - E_p$, where $E_m$ is the total magnetic energy in a volume V, given by
\begin{equation}
    E_m = \frac{1}{2\mu_0} \int_{V} \bm{B}^2 dV ~~,
\label{equ2}
\end{equation}
and $E_p$ is the magnetic energy of the corresponding potential field $\bm{B}_p$, given by
\begin{equation}
	E_p = \frac{1}{2\mu_0} \int_{V} {\bm{B}_p}^2 dV ~~.
\label{equ3}
\end{equation}

To calculate the values of magnetic helicity and magnetic energy, the finite volume method (FV) is often used \citep{Thalmann2011, Valori2012, Yang2013, Amari2013, Rudenko2014, Moraitis2014}. In principle, using Equation (1) one can do an integration to get the relative magnetic helicity numerically with any gauge, as long as the same gauge is used for calculating $\bm{A}$ and $\bm{A}_p$. In this paper, we adopt a finite volume method based on the Coulomb gauge \citep{Yang2013, Yang2018} to calculate the magnetic helicity and magnetic free energy of the coronal field\footnote{\url{https://sun.bao.ac.cn/NAOCHSOS/rmhcs.htm/}}.

\subsection{The flux of magnetic helicity and energy through photosphere}

The relative magnetic helicity and magnetic energy in a volume could be affected by the flux of magnetic helicity and energy transferred through boundaries. It is generally believed that, among all six boundaries, the helicity and energy flux through the bottom boundary (that is, the photosphere) is  the most significant. The helicity flux through the photosphere can be estimated by using the measured photospheric magnetic field. We used the formula given by \citet{Berger1984b}, that is, 
\begin{equation}
\frac{dH_R}{dt}= 2 \int_{S}(\bm{A}_p \cdot \bm{B}_t)V_n dS - 2 \int_{S}(\bm{A}_p \cdot \bm{V}_t)B_n dS \ ,
\label{equ4}
\end{equation}
to estimate the magnetic helicity flux. Here the subscripts of ``$t$'' and ``$n$'' denote the tangential and normal components, respectively, for both the vector magnetic field and the velocity field. $\bm{A}_p$ is the vector potential of the current-free field $\bm{B}_p$, as the one in Equation (1). 

Note that to use this formula to calculate the magnetic helicity flux, $\bm{A}_p$ needs to obey the Coulomb gauge, that is, 
\begin{equation}
\left.\begin{lgathered}
    \nabla \times \bm{A}_p \cdot \bm{\hat{n}} = B_n   \\
    \nabla \cdot \bm{A}_p = 0  \\
    \bm{A}_p \cdot \bm{\hat{n}} = 0  \ .
\end{lgathered} \right.
\label{equ5}
\end{equation}
Otherwise, as pointed out by \citet{Pariat2015}, more terms need to be taken into account when considering the time variation of relative magnetic helicity.

Similar to the magnetic helicity flux, the magnetic energy flux can be estimated \citep{Kusano2002} using observed photospheric magnetograms by the following formula:
\begin{equation}
\frac{dE_m}{dt}= \frac{1}{4\pi} \int_{S} B_t^2 V_n dS - \frac{1}{4\pi} \int_{S}(\bm{B}_t \cdot \bm{V}_t)B_n dS \ .
\label{equ6}
\end{equation}

In this study, we use an optical flow method, that is, the Differential Affine Velocity Estimator for Vector Magnetograms (DAVE4VM), to derive the vector velocity field in the photosphere \citep{Schuck2008}.  We use the time series of vector magnetograms obtained by HMI/SDO, as described in section 2.2. In using DAVE4VM, window size is the most important parameter. In this paper, we use a window size of 19 pixels, about $9.5''$, a number that has been tested and widely used such as in \citet{Liu2012} and \citet{Song2015}.

\section{Analysis and Results}
\label{sect:Res}

In order to study the changes of magnetic helicity and magnetic free energy during solar flares, we need to estimate the magnetic helicity and magnetic free energy before (termed as $H_R^0$ and $E_f^0$, respectively) and after (termed as $H_R^1$ and $E_f^1$, respectively) a flare event. For this purpose, we download 19 HMI/SDO vector magnetograms for each flare event, with their times of observation evenly spreading around each flare peak time. Since each of these magnetograms is about 12 minutes apart, the time span we studied for each flare is about 216 minutes.

\begin{figure}[!htbp]
   \centering

  \begin{minipage}{\textwidth}
      \centering
     \includegraphics[width=0.95\textwidth, angle=0]{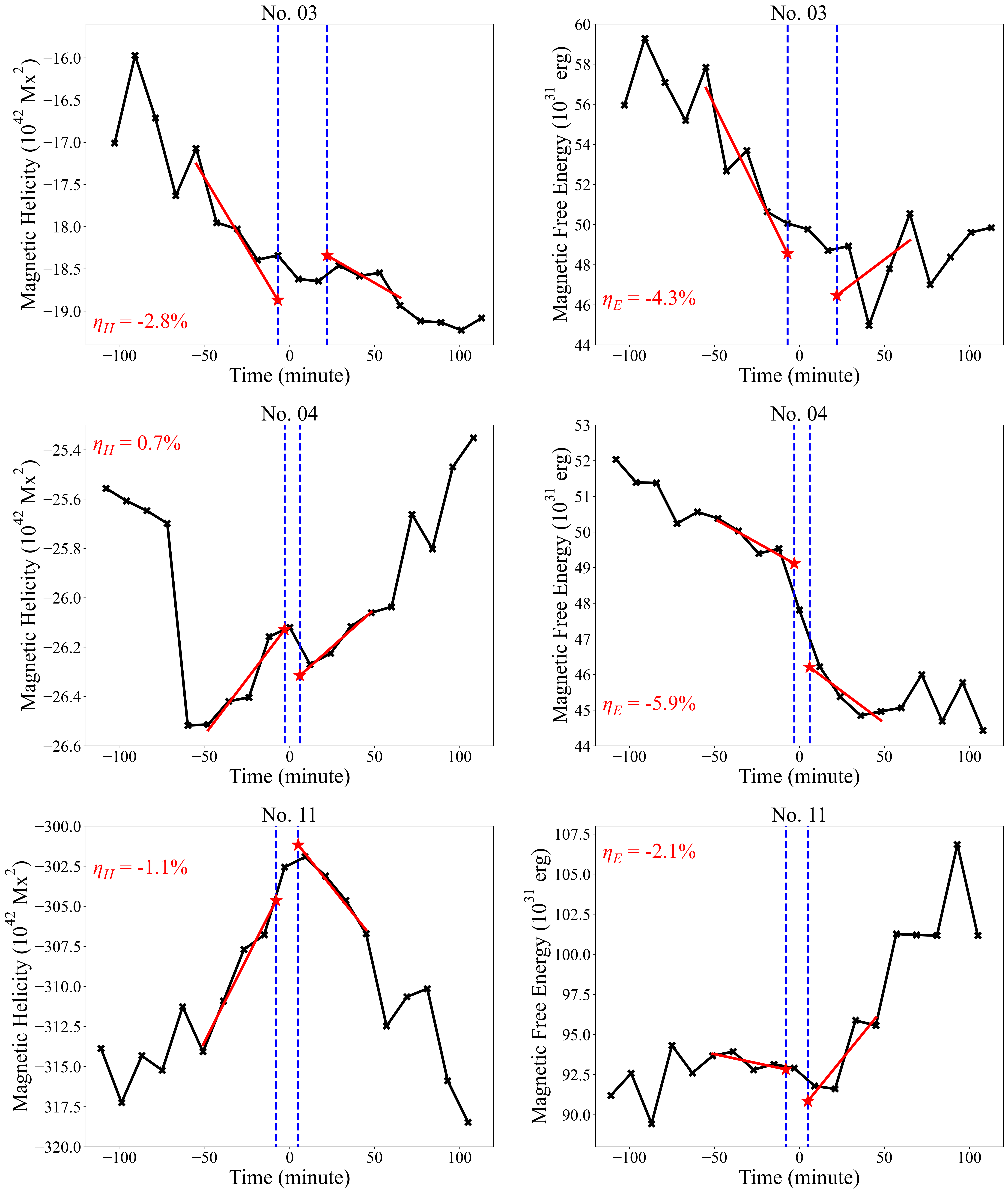}
   \end{minipage}
   \caption{Examples of time profiles of the magnetic helicity (left panels) and magnetic free energy (right panels) of three confined flares (No.03, No.04 and No.11, from top to bottom panels). In each panel, $t=0$ defines the peak time of the flare and the two blue dotted lines show the start and end times of each flare. In each time profile, 4 data points before the flare start time are used for a linear fitting (red line) to get the $H_R^0$ or $E_f^0$. Similarly, the 4 data points after the flare end time is used to get the $H_R^1$ or $E_f^1$. The red stars, intersections of the red lines and the blue lines, in each panel show the obtained values. The change ratios are also shown in each panel in red letters. See text for more details.}
   \label{Fig1}
\end{figure}

\begin{figure}[!htbp]
   \centering
   \begin{minipage}{\textwidth}
      \centering
       \includegraphics[width=0.95\textwidth, angle=0]{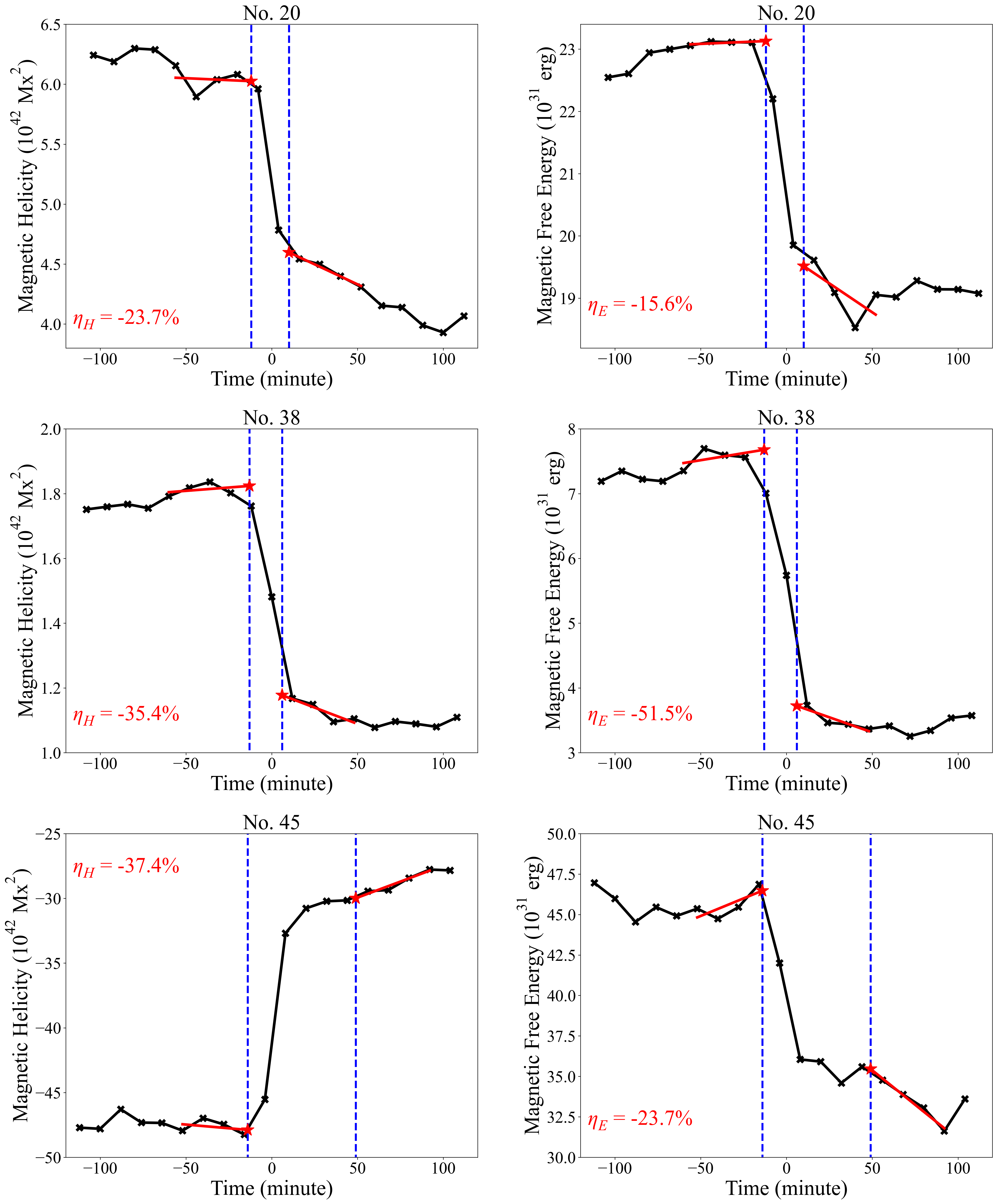}
   \end{minipage}
   \caption{Same as Figure \ref{Fig1} but for eruptive flares, that is, No.20, No.38 and No.45 flares, from top to bottom panels.}
   \label{Fig2}
\end{figure}

We then do a NLFFF extrapolation using each of these magnetograms and based on the extrapolated field we calculate the magnetic helicity and magnetic free energy using the methods and formula presented in the previous section. Totally we have done $19\times47=893$ extrapolations as well as corresponding FV calculations. This gives us 47 time profiles of magnetic helicity evolution and 47 time profiles of magnetic free energy evolution, some of them are presented in Figures \ref{Fig1} and \ref{Fig2} as examples.

Before using these time profiles to estimate the $H_R^0$, $E_f^0$, $H_R^1$ and $E_f^1$ values, we first checked how good our extrapolations are. There are two controlling parameters that are widely used in the community to quantify the quality of field extrapolation. They are: the volume-averaged fractional flux, $f_i$, which quantifies the divergence of the NLFFF model solutions;  the current-weighted average of the angle between the magnetic field and electric current density, $\theta_J$ (CWtheta), which tests the consistency and reliability of the force-free model solutions \citep{Wheatland2000,Schrijver2006}. These two parameters are given by,
\begin{equation}
\left.\begin{lgathered}
    f_i = \frac{\int_{\Delta S_i} \bm{B} \cdot \bm{dS}}{\int_{\Delta S_i} \left|B\right| \cdot \bm{dS}} \approx \frac{(\nabla \cdot \bm{B})_i \Delta V_i}{B_i A_i} = \frac{\nabla \cdot \bm(B_i)}{6\left|\bm{B}\right|_i / \Delta}  \\
    \theta_J = \arcsin \sigma_J = \arcsin [(\sum_i \frac{\left| \bm{J_i} \times \bm{B_i} \right|}{B_i} ) / \sum_i J_i] \ .
\end{lgathered} \right.
\label{equ7}
\end{equation}
where $\Delta$ is the uniform grid size, $A_i$ is the surface area of the grid volume $i$, and $\bm{J_i}$ is the current density of the magnetic field $\bm{B_i}$ in the grid volume $i$.

\begin{figure}[!htbp]
   \centering
       \includegraphics[width=0.95\textwidth, angle=0]{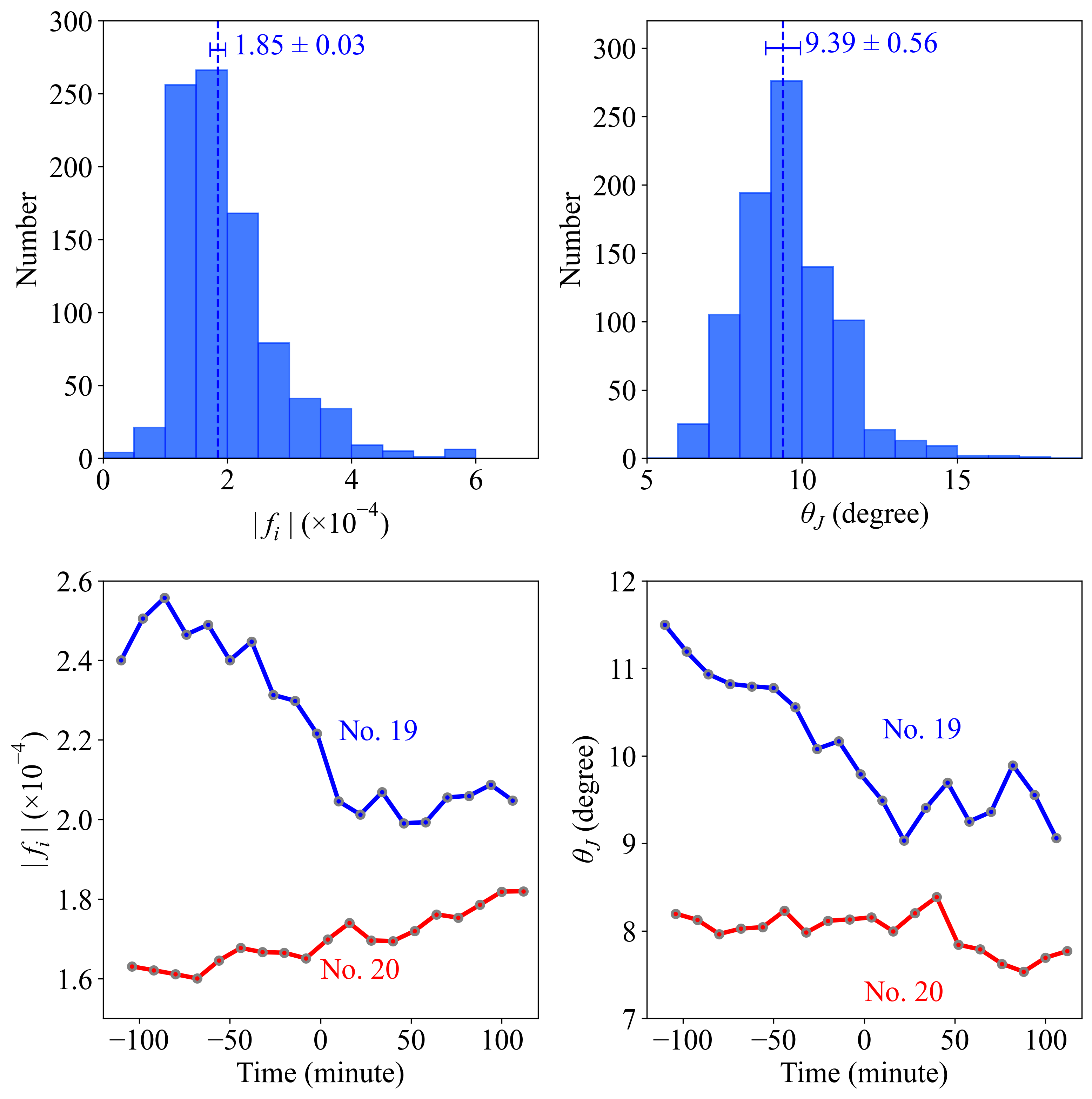}
   \caption{The upper left and right panels show the histograms of $|f_i|$ and $\theta_J$, respectively. The lower left and right panels present the time evolutions of $|f_i|$ and $\theta_J$, respectively, for No.19 (in blue) and No.20 (in red) flares.}
 \label{Fig3}
\end{figure}

We calculate these two parameters for each of our 893 extrapolated fields, and the histograms of them are presented in the upper panels of Figure \ref{Fig3}. The median values of $|f_i|$ and $\theta_J$ are $1.85 \times 10^{-4}$ and 9.39 degrees, respectively. These numbers are pretty low compared to those in most of previous studies. For example, \citet{Thalmann2019b} studied the reliability of magnetic energy and helicity computations based on NLFFF models. They did various extrapolations based on the same magnetogram and presented a good (their Series II) and a bad (their Series I) case in their Figure 2. Since the active region NOAA 11158 they studied is also in our sample (the No.19 and No.20 flares), we plot the time profiles of our $|f_i|$ and $\theta_J$ values in the bottom panels of Figure \ref{Fig3}. The median values of $|f_i|$ and $\theta_J$ are $2.2\times10^{-4}$ and 9.89 degrees, respectively, for the No.19 flare, and are $1.7\times10^{-4}$ and 8.03 degrees, respectively, for the No.20 flare. Comparing ours with theirs in their Figure 2, we see that our $|f_i|$ and $\theta_J$ values are similar to and even a little better than those in their good one (their Series II) case. Actually the helicity values we obtained are also close to those in their good one case, as presented in their Figure 3, and are very different from those in their bad one case.

Now we come to estimate the magnetic helicity and magnetic free energy before ($H_R^0$ and $E_f^0$, respectively) and after ($H_R^1$ and $E_f^1$, respectively) a flare event. Examples are given in Figures 1 and 2.  In these figures, the left panels present the time evolution profiles of relative magnetic helicity and the right panels of magnetic free energy. The three rows in Figure \ref{Fig1} give three examples of confined flares (No.03, No.04 and No.11 from top to bottom), and Figure \ref{Fig2} give three examples of eruptive flares (No.20, No.38 and No.45 from top to bottom). 

In each panel of Figures \ref{Fig1} and \ref{Fig2}, $t=0$ defines the peak time of the flare. The two blue dotted lines show the start and end times of each flare. In each time evolution profile, we use the four data points before the flare start time to do a linear fitting to get the $H_R^0$ or $E_f^0$. Similarly, the four data points after the flare end time are used in a linear fitting to get the $H_R^1$ or $E_f^1$. The red stars in each panel present the intersections of the fitting lines (red lines) and the blue dashed lines, which give the values of $H_R^0$ and $H_R^1$, respectively, in the left panels, or $E_f^0$ and $E_f^1$, respectively, in the right panels. 

Take the upper left panel of Figure \ref{Fig1} as an example. This is the No.3 flare in Table 1. The start time is $-7$min before the flare peak time. The four data points before $t=-7$min are at $t=-55, -43, -31, -19$min. Using $H_R$ values at these four time points, a linear fitting gives $H_R=-3.36\times10^{40}\times t-1.91\times10^{43}$. Extrapolating this fitting line to $t=-7$min, we get  $H_R^0=-1.887 \times 10^{43}$ Mx$^2$. Similarly, the end time of this flare is at $t=22$min. The four data points after the flare end time are at $t=29, 41, 53, 65$min. A linear fitting of $H_R$ at these four points gives $H_R = -1.16\times10^{40} \times t - 1.81\times10^{43}$. Extrapolating this fitting line to $t=22$ min gives  $H_R^1=-1.834 \times 10^{43}$ Mx$^2$. 

Similar processes are used to get the $E_f^0$ and $E_f^1$ values. And all these numbers, $H_R^0$, $H_R^1$, $E_f^0$ and $E_f^1$, are listed in Tables 1 and 2.

For each linear fitting we have also calculated the standard deviation, as the measurement error of the fitting. We still use the No.03 flare as an example. The values of relative magnetic helicity $H_R$ at the four data points before the flare are $-1.707 \times 10^{43}$ Mx$^2$, $-1.795 \times 10^{43}$ Mx$^2$, $-1.803 \times 10^{43}$ Mx$^2$ and $-1.839 \times 10^{43}$ Mx$^2$. Putting the times of these data points, that is, $t=-55, -43, -31, -19$min, into the linear fitting line we get $H_R^{fit}=-1.725 \times 10^{43}$ Mx$^2$, $-1.766 \times 10^{43}$ Mx$^2$, $-1.806 \times 10^{43}$ Mx$^2$ and $-1.847 \times 10^{43}$ Mx$^2$. Then the error of $H_R^0$ for No.03 flare is derived as
\begin{equation}
    \sigma_H = \sqrt{ \frac{\sum_{i=1}^{4} [H_R(t_i)-H_R^{fit}(t_i)]^2}{4}} \ ,
\label{equ8}
\end{equation}
which gives a number of $1.8\times10^{41}$ Mx$^2$. The errors of $H_R^1$,  $E_f^0$ and $E_f^1$ are obtained in a similar way. Values of these estimated errors are also presented in Tables \ref{Tab1} and \ref{Tab2}, as $-18.87\pm 0.18$ ($\times 10^{42}$ Mx$^2$) for $H_R^0$ of No.03 flare for example. 

\begin{figure}[!htbp]
   \centering
   \begin{minipage}{\textwidth}
       \centering
       \includegraphics[width=0.48\textwidth, angle=0]{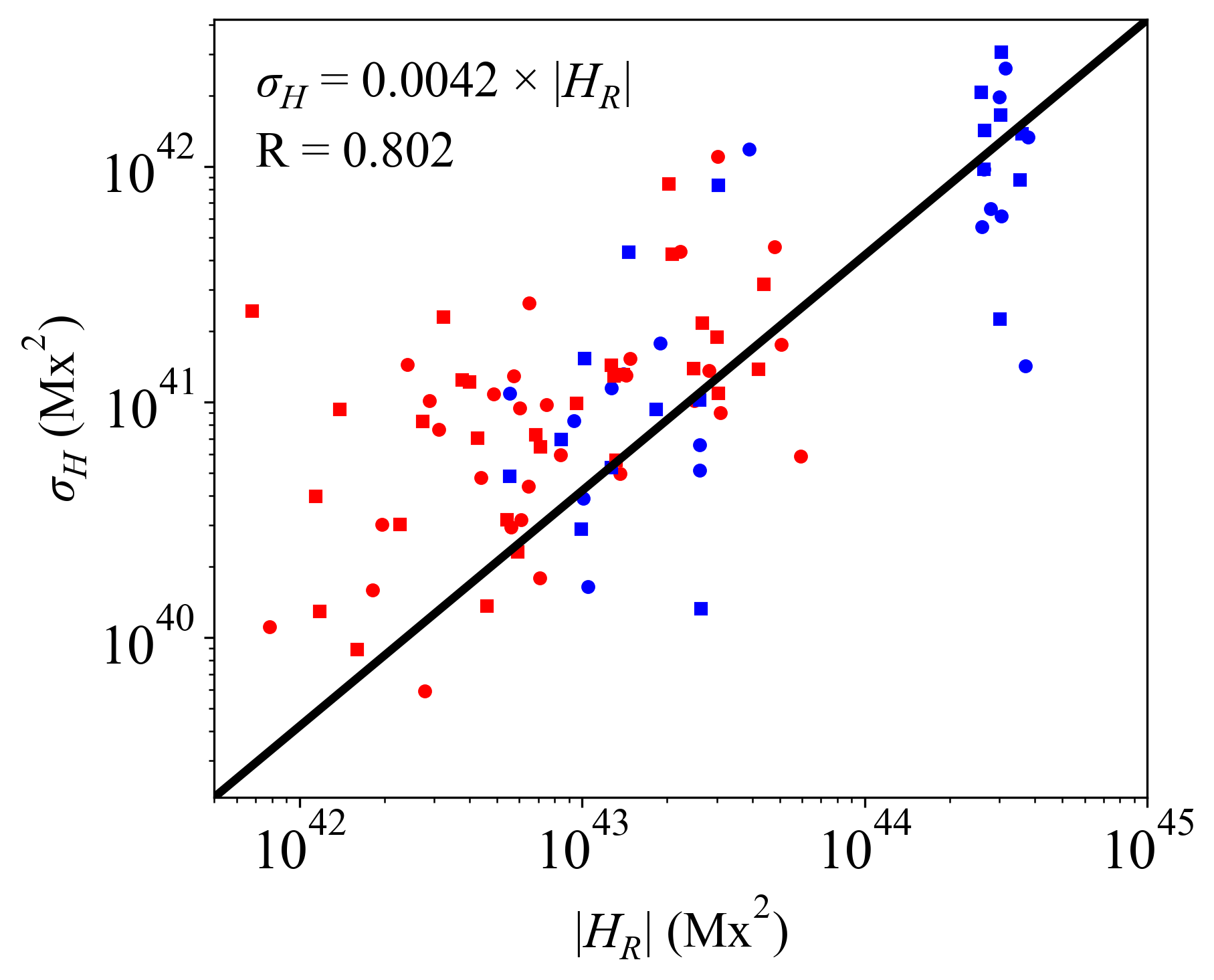}
       \hspace*{2mm}
       \includegraphics[width=0.48\textwidth, angle=0]{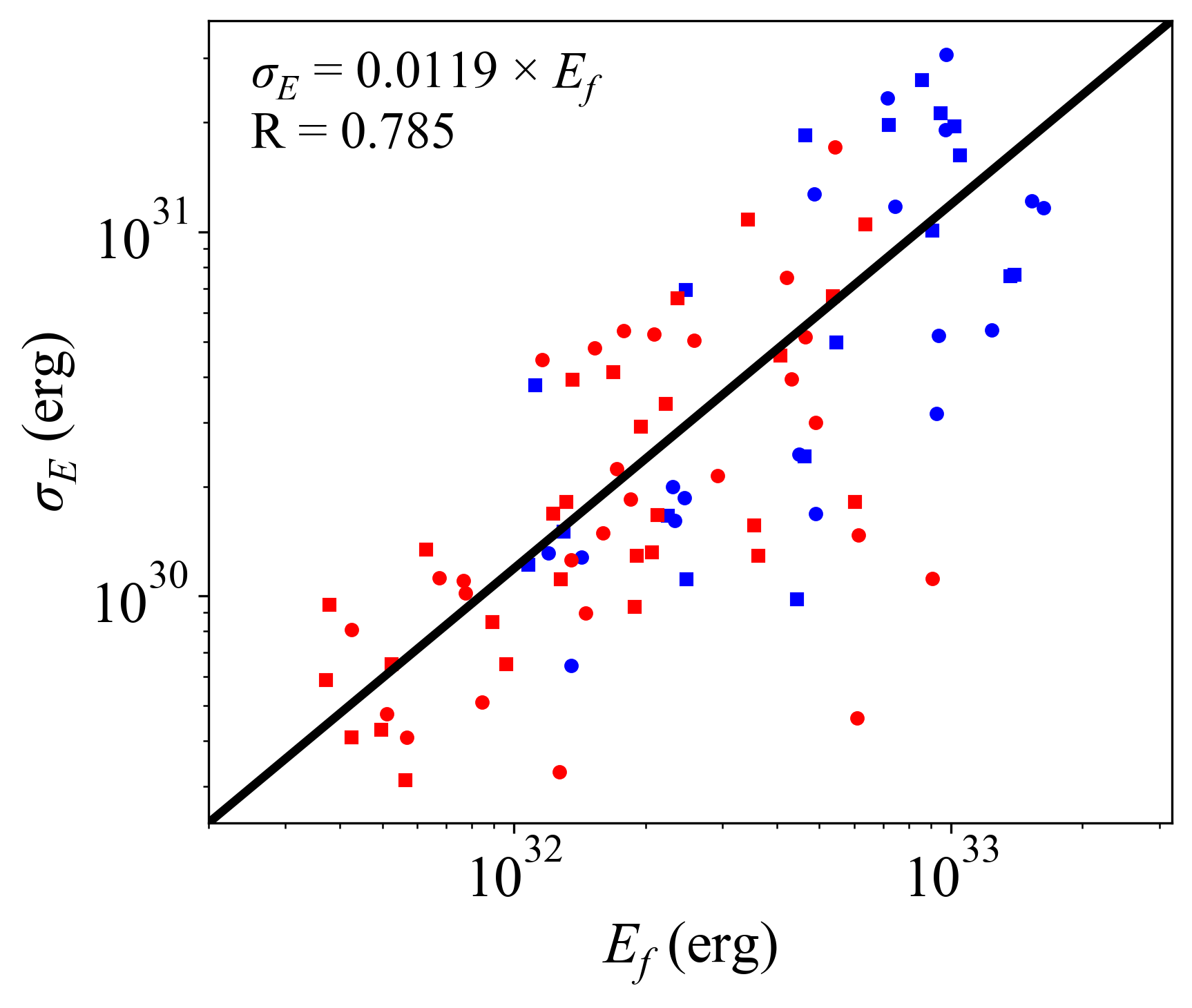}
   \end{minipage}
   \caption{Plots of estimated errors ($\sigma_H$ or $\sigma_E$) against their measured values ($|H_R|$ or $E_f$). The blue and red points are for confined and eruptive flares, respectively. The ``X'' and ``O" symbols are of before and after flares, respectively. A linear fitting (black line) has been given between each x- and y- values,  result of which is listed at the upper left corner of each panel.}
 \label{Fig4}
\end{figure}

\begin{figure}[!htbp]
   \centering
   \begin{minipage}{\textwidth}
       \centering
       \includegraphics[width=0.95\textwidth, angle=0]{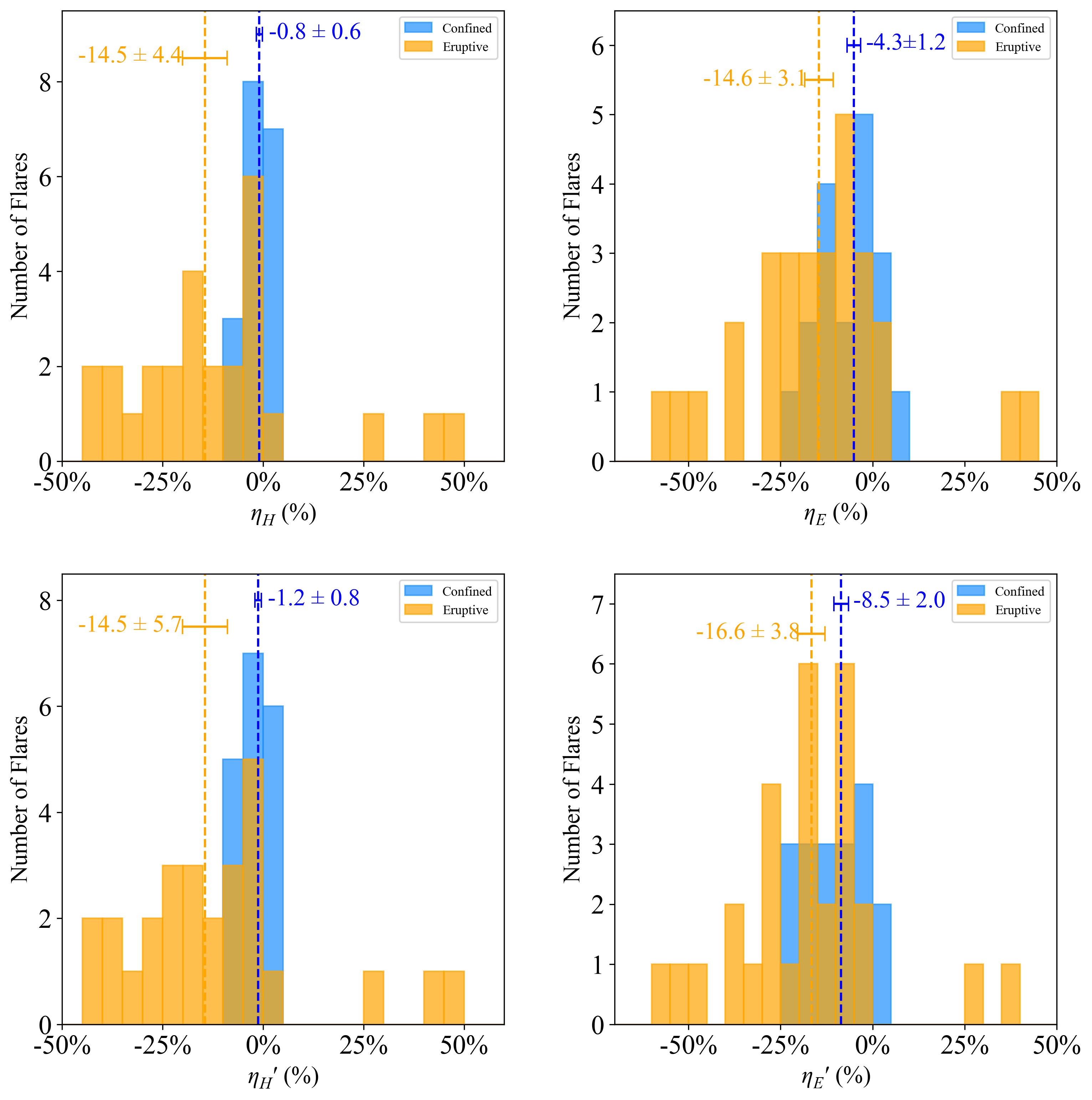}
   \end{minipage}
   \caption{Histograms of the change ratios of magnetic helicity (left panels) and magnetic free energy (right panels) without (top panels) or with (bottom panels) photospheric transferred flux considered. The confined flares and eruptive flares are shown in blue and orange, respectively, with the dotted lines showing their median values. Standard deviation errors are also estimated and presented.}
 \label{Fig5}
\end{figure}

In Figure \ref{Fig4} we present these measurement errors against their measurement values. Left panel is for the helicity measurement and the right panel for magnetic free energy.  The blue and red points in the figure represent the confined flares and eruptive flares, respectively. The ``X'' and ``O" symbols indicate whether the values are of before and after flares, respectively. A linear fitting between the errors of magnetic helicity ($\sigma_H$) and the magnitudes of helicity ($|H_R|$) gives $\sigma_H=0.0042|H_R|$, which means that the measurement errors are about 0.42\%, on average, of their values for magnetic helicity. Similarly, a linear fitting between the errors of magnetic free energy ($\sigma_E$) and the magnitudes of magnetic free energy ($E_f$) gives $\sigma_E=0.0119 E_f$, which means that the measurement errors are about 1.2\%, on average, of their values for magnetic free energy. These numbers are smaller than the change ratio values that we are going to see in Figure \ref{Fig5}, which indicates that the change ratios we obtained are not caused by measurement errors. 

Also listed in Tables 1 and 2 and presented in Figures 1 and 2 are the change ratios of the magnetic helicity ($\eta_H$) and the magnetic free energy ($\eta_E$). They are defined as
\begin{equation}
    \eta_H =  \frac{H_R^1 - H_R^0}{H_R^0}  ~~~~, 
   \hspace{1cm}  \eta_E =  \frac{E_f^1 - E_f^0}{E_f^0}  \ .
\label{Equ9}
\end{equation}


Using Equations (4) and (6), we have also calculated the magnetic helicity and magnetic energy transfer fluxes through the photosphere during these flare times. We integrate the obtained transfer fluxes from $t_{start}$ to $t_{end}$ to get an estimation of the magnetic helicity and magnetic energy transferred into the corona through the photosphere during the flare time, that is,
\begin{equation}
\left.\begin{lgathered}
    dH_R =  \int_{t_{start}}^{t_{end}} \frac{dH_R}{dt} dt \\
    dE_m =  \int_{t_{start}}^{t_{end}} \frac{dE_m}{dt} dt \ .
\end{lgathered} \right.
\label{Equ10}
\end{equation}
The values of these transferred helicity ($dH_R$) and energy ($dE_m$) are also listed in Tables \ref{Tab1} and \ref{Tab2}. We can see that they are significantly smaller than the coronal content, a point that has also been found in \citet{Liu2023}. 

To account for these transferred values, a new set of change ratios of magnetic helicity ($\eta_H^{\prime}$) and magnetic free energy ($\eta_E^{\prime}$) has been calculated,  defined as 
\begin{equation}
    \eta_H^{\prime} =  \frac{H_R^1 - H_R^0 - dH_R}{H_R^0}  ~~~~,
  \hspace{1cm}  \eta_E^{\prime} =  \frac{E_f^1 - E_f^0 - dE_m}{E_f^0}  \ .
\label{Equ11}
\end{equation}
These values, $\eta_H^{\prime}$ and $\eta_E^{\prime}$, have also been listed in Tables \ref{Tab1} and \ref{Tab2}. 

From Figures \ref{Fig1} and \ref{Fig2} we can see that, the magnitudes of $\eta_H$ and $\eta_E$ are both relatively small for the confined flares, and in contrast, the magnitudes of the $\eta_H$ and $\eta_E$ of the eruptive flares are significantly larger than those of the confined flares. The $\eta_H$ and $\eta_E$ are $-2.8\%$  to $0.7\%$ and $-2.1\%$ to $-5.9\%$, respectively, for the three confined flares, whereas the $\eta_H$ and $\eta_E$ are $-23.7\%$ to $-37.4\%$ and $-15.6\%$ to $-51.5\%$, respectively, for the three eruptive flares. The relatively small values of $\eta_H$ of the confined flares verify the conservation property of the magnetic helicity during the fast magnetic reconnection, whereas the relatively large values of $\eta_H$ of the eruptive flares indicate that the magnetic helicity and energy have been carried away by CMEs. 

This is more evident in Figure \ref{Fig5}, where the histograms and the median values of the change ratios of both magnetic helicity and magnetic free energy are presented, with confined flares in blue and eruptive flares in orange. It is obvious that the distributions of $\eta_H$ and $\eta_E$, as well as  $\eta_H^{\prime}$ and $\eta_E^{\prime}$, of eruptive flares are wider than those of the confined flares. The median values of $\eta_H$ and $\eta_H^{\prime}$ are small ($-0.8\%$ and $-1.2\%$ respectively) for confined flares, which confirms the conservation property of magnetic helicity. 

The median values of $\eta_E$ and $\eta_E^{\prime}$ of the confined flares are also small ($-4.3\%$ and $-8.5\%$ respectively), but their magnitudes are relatively larger than those of $\eta_H$ and $\eta_H^{\prime}$. This is consistent with the statement by \citet{Berger1984a} that the total magnetic helicity is still conserved during magnetic reconnection even when there is a magnetic energy dissipation. 

Most notably are the median values of the $\eta_H$ and $\eta_H^{\prime}$ of eruptive flares ($-14.5\%$ for both), whose magnitudes are evidently larger than those of the confined flares ($-0.8\%$ and $-1.2\%$ respectively). This supports the statement that CMEs take away magnetic helicity \citep{Low1996}. 

It is also interesting to notice that, the magnitudes of the median values of $\eta_E$ and $\eta_E^{\prime}$ of eruptive flares ($-14.6\%$ and $-16.6\%$ respectively) are also larger than those of the confined flares ($-4.3\%$ and $-8.5\%$ respectively). This suggests that, in addition to providing the thermal energy as that in a confined flare, an eruptive flare needs to consume more magnetic free energy in order to drive the plasma of the CME. This is also a picture that is consistent with our theoretical understandings, but it is consoling that our calculations using observational data are able to detect the differences.

\section{Summary and Discussion}
\label{sect:Sum}

In this study we construct a sample of 47 major solar flare events in the $24^{th}$ solar cycle, which contains 18 no-CME-associated confined flares and 29 CME-associated eruptive flares. We apply the NLFFF extrapolations to a series of HMI/SDO vector magnetograms and use FV methods to derive the magnetic helicity and magnetic free energy of these flares. We find that the confined flares and the eruptive flares distinguish themselves from each other in the statistics. The median values of the change ratios of both magnetic helicity and magnetic free energy show significant larger values in magnitude for the eruptive flares, whereas the corresponding numbers are relatively small for the confined flares. These results, using observational data, confirm following theoretical understandings: (1) The total magnetic helicity is approximately conserved during magnetic reconnection; (2) The total magnetic helicity is still conserved even when there is a magnetic energy dissipation; (3) CMEs take away magnetic helicity; (4) Eruptive flares consume more magnetic energy than confined flares in order to drive the CME plasma.

It is interesting to notice that \citet{Liu2023} also studied the evolution of magnetic helicity and energy associated with major solar flares. Their sample consists of 21 X-class flares from 2010 to 2017, a little bit smaller than ours. Since their main purpose is different from ours, their temporal profile has been obtained by averaging all studied flares. However, it is consoling that their data also present the same tendency which indicates ``the systematical removal of magnetic helicity from the corona by jettisoned coronal ejecta'' and that ``these flares are an insignificant sink for helicity''.

It is also worthy of mentioning that, in our sample, there are three eruptive flares (No.26, No.33 and No.43) that show large positive $\eta_H$ and $\eta_H^{\prime}$ values. We do not think these are caused by measurement errors. We checked the AIA images of these three active regions and found that they have a common feature: some part of the field are connected to neighborhood fields outside our calculation box. The neighborhood could be a nearby prominence (as in No.43 flare case) or nearby active regions. We speculate that magnetic helicity exchange between the neighboring active regions or prominences has happened (see \citet{Yang2009} for an example), resulting in an ``appearing'' large positive helicity change ratio in these three cases. Magnetic energy may also be transferred into these regions through side boundaries. It is worthy of checking these possibilities by studying the magnetic helicity and energy evolutions in a larger box than that in current study and see whether there is really a helicity and energy transfer between the currently studied active region and the neighborhood.  It would be interesting to check all these in future studies.

\acknowledgements
We thank HMI team for providing excellent magnetogram data online and the anonymous referee for helpful comments and suggestions. This study is supported by the National Natural Science Foundation of China (grant Nos. 12250005, 12073040, 11973056, 12003051, 11573037, 12073041, 11427901, 11973056, 12173049 and 11611530679); the National Key R\&D Program of China (grant No. 2022YFF0503800, 2021YFA1600503); the Strategic Priority Research Program of Chinese Academy of Sciences (grant Nos. XDB09040200, XDA15010700 and XDA15320102); by the Youth Innovation Promotion Association of CAS (2019059); and the ISSI International Team on Magnetic Helicity estimations in models and observations of the solar magnetic field. This research is also supported by the mobility program (M-0068) of the Sino-German Science Center, the International Space Science Institute (ISSI) in Bern and the ISSI-BJ in Beijing, through ISSI International Team project 568 and ISSI-BJ International Team project 55 (Magnetohydrostatic Modeling of the Solar Atmosphere with New Datasets).






\end{document}